\documentclass[12pt]{article}
\usepackage{amsmath,amssymb,amsfonts,amsbsy}
\usepackage{graphicx}
\newcommand{\ba}{\begin{eqnarray}}
\newcommand{\ea}{\end{eqnarray}}
\newcommand{\beqs}{\begin{eqnarray}}
\newcommand{\eeqs}{\end{eqnarray}}

\begin{document}
\setcounter{section}{0}
\setcounter{subsection}{0}
\setcounter{equation}{0}
\setcounter{figure}{0}
\setcounter{footnote}{0}
\setcounter{table}{0}
\begin{center}
\textbf{The slope of the hadron spin-flip amplitude  and
 the determination of $\rho(s,t)$
 }\bigskip

 \underline{O.V. Selyugin}\footnote{selugin@theor.jinr.ru}, J.-R. Cudell\footnote{jr.cudell@ulg.ac.be},
E. Predazzi\footnote{predazzi@to.infn.it}\bigskip

{\small
  $^1$\it{BLTPh, Joint Institute for Nuclear Research, Dubna, Russia} \\
$^{\,2}$\it IFPA, AGO Dept., Universit\'e de Li\`ege, Li\`ege, Belgium\\
$^{\,3}$ \it Dipartamento di Fisica Teorica, Universit\`a di Torino, Italy\\
}

\end{center}
\vspace{0.0mm} 
\begin{abstract}
 We re-examine the extraction of  $\rho(s,t)$, the ratio of the real part to the
 imaginary part
 of the scattering amplitude, and of the spin-flip amplitude, from the existing
 experimental data in the Coulomb-hadron interference region. We show that it is not
 possible to find reasonable assumptions
 about the structure of the scattering amplitude of proton-proton and proton-antiproton
 elastic scattering at high energy
 that would lead, in proton-antiproton scattering
 for $3.8 < p_L <6.0$ GeV/c,  to an agreement between data and an analysis
   based on dispersion relations.
\end{abstract}
\vspace{7.2mm}
\section{Introduction}

The calculation  via dispersion relations of the ratio of the real part to the imaginary
part of the forward spin-non-flip amplitude, $\rho(s,t) $, does not agree with the data
until one gets to high energies, and it misses all the interesting intermediate-energy
structures.

On the theory side, the situation is very complex and uncertain. Analyticity showed that
one could not do without a real part, while
polarization data proved that it was not possible to ignore spin complications, as
the real part of the spin-non-flip amplitude has a zero, around which the contribution
of the spin-flip amplitude, which decreases quite slowly with energy, cannot be ignored.

On the experimental side, the situation is not entirely clear cut either
\cite{Rev-LHC}, and one of the difficulties is  due to  the
 lack of  experimental data at high energies and small momentum transfer.

 In this talk, we consider in great detail the situation concerning $\rho(s,t)$.
   The model we propose takes into account all known features of the near-forward
   proton-proton and proton-antiproton data, {\it i.e.} different slopes for the
   spin-non-flip  and the spin-flip amplitudes, the value of total cross sections and of
   $\rho(s,t)$, the relative phase of the Coulomb and hadron amplitudes and the
   form factors of the nucleons.

\section{Impact of the Coulomb-hadron phase}
 Let us first compare different approximations
   for the Coulomb-hadron interference used in fits to the experimental
   $p\bar{p}$-scattering data \cite{Disser-data}. First, we use
   the simple West-Yennie form of the relative phase \cite{wy}.
   This leads to values for $\rho(s,0)$ shown in the second column of Table 1.
   The results show the distribution of the values  of $\rho(s)$ extracted
   from the  experiments. In  two cases,  they lie  slightly above
   $\rho_{exp}$ (at $p_L = 4.066, \ 5.603,  \ 5.94 \ $ GeV/c); in three cases
   they lie considerably higher than $ \ \rho_{exp}\ $ (at $p_L = 5.72, \ 6.23 \ $
   GeV/c)   and in one case they lie below (at $p_L = 3.7 \ $GeV/c).

   If we take the slightly more complicated phase proposed by Cahn \cite{can},
   the results are almost the same (see the third column of Table 1).
   Finally, if we use the expression derived by one of us \cite{selmp1,PRD-Sum},
   taking into account the two-photon amplitude and using a dipole form factor,
   the fit gives different values for $\rho(s)$   (see the last column  of Table 1):
   the results lie above $\rho_{exp}$ for all the considered energies, so that
   the difference with the predictions of the dispersion analysis gets worse, as shown in Fig. 1.
\label{sec:figures}
\begin{figure}
\begin{center}
\includegraphics[width=0.45\textwidth] {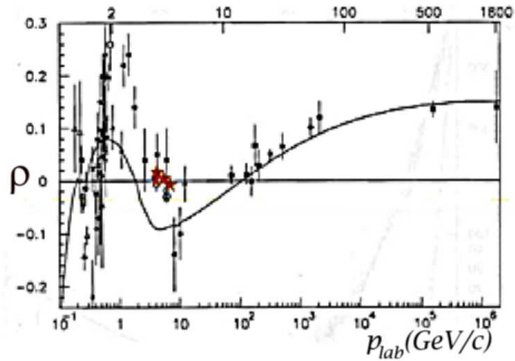}
\end{center}
 \caption{$\rho(s,0)$ - the ratio of the real part to the imaginary part of the
 elastic scattering amplitude for proton-antiproton scattering at low
 energies. The curve shows the dispersion relation description for
  $p \bar{p}$ scattering \cite{Kroll}, and the stars are the result of our analysis.
 } \label{Fig_2}
\end{figure}
\begin{table*}
 \caption{The dependence of $\rho(s,0)$ on the model used for the Coulomb-hadron phase in proton-antiproton scattering. N is the number of data points.} \label{Table-3}
\vspace{.1cm}
\begin{center}
\begin{tabular}{|c|c|c||c|c|c|} \hline
              &               &               &        &    &       \\
\large{ $p_{L}(GeV/c)$ }& \large{N} &\large{ $\rho_{exper.}$ }\cite{Disser-data}& \large{
$\rho$}(${phase \cite{wy}})$  &
 \large{$\rho$}($ {phase \cite{can} }$)& \large{$\rho$}(${phase \cite{selmp1,PRD-Sum}})$ \\
              &               &               &        &    &       \\
 \hline
              &               &               &        &    &       \\
3.702 & 34 & $+0.018 \pm 0.03$ &  $+0.0077 \pm 0.02$  &  $+0.0078 \pm 0.08$ & $+0.028
\pm 0.08$\\
4.066 & 34 & $-0.015 \pm 0.03$ &  $+0.0377 \pm 0.02$  &  $+0.0378 \pm 0.08$ & $+0.0324
\pm 0.08$\\
5.603 & 215&  $-0.047 \pm 0.03$ &  $+0.035 \pm 0.02$  &  $+0.036 \pm 0.08$ & $-0.0017
\pm 0.08$\\
5.724 & 115&  $-0.051 \pm 0.03$ &  $+0.0139 \pm 0.02$  &  $+0.014 \pm 0.08$ & $-0.0088
\pm 0.08$\\
5.941 & 140&  $-0.063 \pm 0.03$ &  $-0.0003 \pm 0.02$  &  $-0.004 \pm 0.08$ & $-0.0055
\pm 0.08$\\
6.234 & 34 &  $-0.06 \pm 0.03$ &  $+0.0162 \pm 0.02$  &  $+0.0162 \pm 0.08$ & $-0.0216
\pm 0.08$ \\
              &               &               &        &    &       \\
 \hline
\end{tabular}
\end{center}
 \end{table*}

\section{Impact of the spin-flip amplitude}

    In most analyses, one assumes that
  the imaginary and real parts of the spin-non-flip
  amplitude have an exponential behaviour with the same $t$ slope, and that the
  imaginary and real parts of the spin-flip amplitudes, without the
  kinematic factor $\sqrt{|t|}$, are proportional to the corresponding spin-non-flip parts of the amplitude, with a proportionality constant independent of $s$.
 In \cite{slope-MPL99} it was shown that if the slope of the spin-flip amplitude is bigger than that for spin non-flip, $B_{sf}=2B_{nf}$,
  the contribution of the spin-flip amplitude
  can be felt in the differential
 cross sections of elastic hadron scattering at small $|t|$.
As it is not possible to calculate the hadronic  amplitudes
  from  first principles,
  we have to resort to some assumptions about their $s$ and $t$  dependencies  \cite{PS-EPA02,CPS-EPA04}.

   Here, we use this simple model for the spin-flip
   amplitude and study its impact on the determination of $\rho(s,t)$.
We take
   the spin-non-flip and spin-flip amplitudes in the simplest exponential form
    \ba
       F^{h}_{nf}&=& h_{nf} \ [i+\rho(s,0)] \  e^{B_{nf} t/2};\\
       F^{h}_{sf}&=& \sqrt{-t}/m_{p} \ h_{sf} \ [i+\rho(s,0)] \  e^{B^{sf} t/2},
\ea
with $B_{sf}=2B_{nf}$.
   The differential cross section in this case will be
  \begin{eqnarray}
  \frac{d\sigma}{dt} =
 2 \pi \ [|F_{nf}|^{2} +  2 |F_{sf}|^{2}
  ],    \label{dsdt-sf}
\end{eqnarray}
   where the amplitudes $F_{nf}$ and $F_{sf}$ will include the
   corresponding  electromagnetic parts and the Coulomb-hadron
   phase factors as mentioned previously.

  The results of our new fits of the proton-antiproton experimental data
  for $ p_L$ in  $[3.7,6.2] \ $GeV/c are presented in Table 2.
  The changes of $\chi^2$ after the inclusion
  of the spin-flip amplitude are measured by the coefficient
\ba
 R_{\chi} \ = \ \frac{\chi^2_{ \ without  \ sf.}
 - \chi^2_{ \ with  \ sf.} }{\chi^2_{ \ without \ sf.}}.
  \label{Rchi}
\ea

  We again obtain values of  $\rho$ close to zero and prevalently  positive.
  Once again, as seen from Fig. 1, the results do not agree with the  prediction by the dispersion analysis \cite{Kroll}.

\begin{table*}
 \caption{Spin dependence of proton-antiproton elastic scattering}
\label{Table-3}
\vspace{.1cm}
\begin{center}
\begin{tabular}{|c|c|c||c|c|c|} \hline
              &               &               &          &    & \\
\large{ $p_{L(GeV/c)}$ }  & \large{  N} &\large{ $\rho_{exp.}$ } &\large{$R_{\chi}$}
 &\large{  $\rho_{model}$} &\large{ $h_{sf}$, GeV } \\
              &               &               &          &    & \\ \hline
                  &               &               &          &    & \\
3.702 & 34 & $+0.018 \pm 0.03$ &  $8 \%$  &  $+0.057 \pm 0.02$ & $ 49.8 \pm 1.4$\\
4.066 & 34 & $-0.015 \pm 0.03$ &  $25 \%$  &  $+0.052 \pm 0.009$ & $ 48.9 \pm 0.7$\\
5.603 & 215&  $-0.047 \pm 0.03$ &  $3.5 \%$  &  $+0.014 \pm 0.005$ & $ 35.6 \pm 4.$\\
5.724 & 115&  $-0.051 \pm 0.03$ &  $6.5 \%$  &  $+0.023 \pm 0.004$ & $ 38.2 \pm 4.5$\\
5.941 & 140&  $-0.063 \pm 0.03$ &  $4.5 \%$  &  $+0.007 \pm 0.003$ & $ 43.2 \pm 0.4$\\
              &               &               &          &  &   \\
 \hline
\end{tabular}
\end{center}
   \end{table*}

\section{Conclusion }
    The present analysis, which includes the contributions of Coulomb interference and spin effects,
    shows a  contradiction  between the extracted value of  $\rho(s,0)$
    and the predictions from the analysis based on dispersion relations.

 If such a situation is confirmed by future new data from the LHC experiments, it could
 reveal new
  effects such as , for example, a fundamental length of the order of 1 TeV.

    It is likely, however,  that the theoretical analysis can be further developed, to
    include
     additional corrections connected with possible oscillations in the scattering
     amplitude and
    with the $t$-dependence of the spin-flip scattering amplitude.
     We hope that
    the forward experiments at NICA will also give valuable information for the improvement of our theoretical understanding of this question.

\end{document}